\documentclass[12pt]{article}

\usepackage[english]{babel}
\usepackage[utf8]{inputenc}
\usepackage[T1,T2A]{fontenc}
\usepackage{cite}
\usepackage{physics}
\usepackage{mathtools}
\usepackage{xcolor}
\usepackage{graphicx}
\usepackage{dcolumn}
\usepackage{bm}

\usepackage[utf8]{inputenc}
\usepackage{amsmath}
\usepackage{amssymb}
\usepackage{xcolor}
\usepackage{rotating}
\usepackage{lscape}
\usepackage{longtable}
\usepackage{authblk} 

\usepackage{hyperref}

\usepackage[bottom=0.5in]{geometry}
\usepackage{caption}

\begin{document}

\title{\bf Constraints on Features in the Cosmological Power Spectrum from Observations of the Epoch of Reionization} 
\author[1]{O.R. Skorikov}
\author[1,*]{S.V. Pilipenko}
\author[1]{M.V. Tkachev}
\affil[1]{P.N. Lebedev Physical Institute, Moscow}

\maketitle

\vspace{2mm}

\begin{abstract}
We consider cosmological models with a power spectrum of perturbations featuring an enhanced amplitude on dwarf galaxy scales (with a "bump" or a "tilt"). Early formation of a large number of galaxies in such models, compared to the standard spectrum, can shift the epoch of reionization to higher redshifts compared to observations. We show that for moderate bump amplitude $\mathcal{A}<1.5-2$, the considered models are not ruled out by observations of reionization at $z \approx 8$ due to existing uncertainties in the fraction of ultraviolet photons escaping galaxies, $f_{esc}$, and inhomogeneity of the neutral hydrogen distribution.    
\end{abstract}



\vfill
\noindent\rule{8cm}{1pt}\\
{$^*$ Email: $<$spilipenko@asc.rssi.ru$>$}

\section{Introduction}
\noindent
Although the $\Lambda$CDM cosmological model is well supported by numerous observations and is the simplest, recently it has become popular to consider cosmological models with distorted initial perturbation spectra \cite{Parashari23,Tkachev23,2025PhRvD.112b3512W,Nadler25}. The observational motivation for these works is a possible excess number of galaxies at high redshifts \cite{Naidu2022a,Naidu2022b,Lovell22}. Theoretically, such models have been considered for a long time, see e.g. \cite{Ivanov94}. The shape of the power spectrum is determined by processes during inflation. In the simplest inflation models, the spectrum is a power law with a tilt close to -1. If the inflaton potential has breaks, steps, or if inflation proceeds in multiple stages with different fields, the spectrum may have "bumps", oscillations, or changes in the power-law slope \cite{Inomata23}.

Current observations of large-scale structure and Lyman-alpha forest confirm a power-law initial spectrum up to wavenumbers about $k<1$ Mpc$^{-1}$, corresponding to spatial scales of 6 Mpc \cite{Chabanier19}. Changes in the spectrum on smaller scales up to $k<50$ Mpc$^{-1}$ can have observable consequences, as they affect the halo mass function of dwarf galaxies \cite{Tkachev23,Pilipenko24,Tkachev24,2025PhRvD.112b3512W,Eroshenko25,Dekker25,Nadler25}. These works studied the impact of a bump on the number of bright galaxies and supermassive black holes at high $z$, the number of Milky Way satellite galaxies, density inside dwarf galaxies, dark matter annihilation signals, and the 21 cm line signal from the epoch of galaxy formation onset. Deviations on scales $k>50$ Mpc$^{-1}$ can be constrained by observing distortions of the cosmic microwave background frequency spectrum \cite{Chluba12} and the stochastic gravitational wave background \cite{EPTA24}.

In this work, we examine the influence of power spectrum distortions in the range $k=5-50$ Mpc$^{-1}$ on the production of ionizing Lyman continuum photons at $z>10$. According to \cite{Tkachev23}, distortions in this wavenumber range manifest as a significant increase in the number of galactic-mass halos at $z>10$. At lower redshifts, differences from the power-law spectrum smooth out due to nonlinear large-scale structure evolution. At this epoch, the first galaxies begin to form, and according to observations \cite{Planck18_VI}, about half of all hydrogen in the Universe is reionized by $z \approx 8$. The appearance of many additional galaxies at earlier epochs can alter this picture, allowing constraints on such spectrum distortions from observations.

There are two possible sources of UV photons driving reionization: active galactic nuclei and stars \cite{Fan06}. Current models favor stellar reionization at $z>6$ \cite{Ma15,Barkana16,Madau17}. To compute the number of ionizing photons, we use an approach similar to \cite{Wu24}, constructing cosmological simulations of dark matter halo evolution for different models. A semi-analytical GRUMPY model \cite{Kravtsov22} is applied to estimate stellar population parameters in these halos. Ionizing radiation fluxes are computed using the stellar population synthesis and spectral model BPASS \cite{Eldridge17,Byrne22}.

To ionize hydrogen in the Universe, more UV photons must be emitted into the intergalactic medium within a recombination time than there are hydrogen atoms. The required photon density is calculated in \cite{Madau99}. By comparing the photon densities obtained from the GRUMPY model with the required values, we can find the redshift of reionization for different initial power spectrum models. However, both the galaxy radiation model and the reionization model have uncertainties. The average fraction of UV radiation escaping galaxies ($f_{esc}$) is unknown and often taken as 0.1 but may vary by several times. The number of photons needed for ionization also depends on the clumping factor $C \equiv \langle n_H^2 \rangle / \langle n_H \rangle^2$, which can range from 1 to 30 depending on ionization degree and large-scale structure development. In this paper, we investigate whether any initial spectrum models can be ruled out despite these uncertainties and compare models with each other.

\section{Models and Assumptions}
\noindent
We consider two types of deviations from the power-law initial power spectrum taken from \cite{Tkachev23,Tkachev24}: spectra with a bump and with a tilt. The bump is described by
\begin{equation}
		\frac{P_\mathrm{bump}(k)}{P_\mathrm{\Lambda CDM}(k)} = 1 + A \cdot \exp\left(-\frac{(\log(k) - \log(k_0))^2}{\sigma_k^2}\right),
		\label{bump}
\end{equation}
where $A$ is the dimensionless amplitude, $k_0$ is the bump center wavenumber, and $\sigma_k$ is the dimensionless width.

Note that although formula (\ref{bump}) has three parameters, for narrow bumps $\sigma_k \ll 1$, the halo mass function and evolution depend mainly on the product $A \sigma_k$. This follows from Press-Schechter formalism and is noted in \cite{Tkachev23,Eroshenko25}. In our case, the dimensionless amplitude $\mathcal{A} \equiv A \sigma_k = 2$. This amplitude is required to increase the number of galaxies at $z \sim 10$ consistent with JWST data better than $\Lambda$CDM. Values of $k_0$ were taken as 4.7, 10, 20, 34, 54 Mpc$^{-1}$.

The second type, "tilted" spectrum, is described by
\begin{equation}
    \frac{P_\mathrm{tilt}(k)}{P_\mathrm{\Lambda CDM}(k)} = \sqrt{1 + \frac{1}{p} \left(\frac{k}{k_0}\right)^{2p+2}}.
    \label{tilt}
\end{equation}
At small wavenumbers $k \ll k_0$, this spectrum matches $\Lambda$CDM; at large $k$ it asymptotes to $P_\mathrm{\Lambda CDM}(k) k^{p+1}$. One such model with $k_0=6.8$ Mpc$^{-1}$, $p=0.5$ was considered.

We use dark matter halo simulations previously applied in \cite{Tkachev23,Eroshenko25} with the following parameters: box size 5 Mpc$/h$ ($\approx 7.4$ Mpc), where $h=0.677$ is the dimensionless Hubble parameter. The box is filled uniformly with $512^3 \approx 130 \times 10^6$ test particles. These parameters allow resolving sufficiently small scales (including all dwarf galaxies) corresponding to $k>50$ Mpc$^{-1}$, while at the simulation end at $z=8$ the box scale still obeys linear theory, avoiding artifacts from periodic boundary conditions.

Initial conditions were prepared using the Ginnungagap code \cite{ggp}. Calculations were performed with the publicly available Gadget-2 code \cite{gadget}. Standard results extracted using the Rockstar halo finder algorithm \cite{Behroozi_2013} include halo mass accretion histories (halo mass vs. redshift). All halos reaching mass $3 \times 10^7 M_\odot$ by simulation end were selected (gas in such halos can cool via atomic hydrogen emission and form normal Population II stars). Merger trees were built, and the most massive progenitor at each simulation output was found. The mass evolution of this progenitor was recorded for use in the semi-analytical model.

GRUMPY \cite{Kravtsov22} is a semi-analytical "regulator"-type model using a minimal set of differential equations describing gas and star evolution. It models atomic and molecular gas mass, stellar mass, gas disk size, and metallicity. It accounts for UV background effects (external and stellar) that modify gas accretion rates onto low-mass halos and star formation rates, as well as gas outflows. GRUMPY reproduces various relations for dwarf galaxies from detailed hydrodynamical cosmological models (e.g., stellar mass to halo mass relation) and observations (gas mass to stellar mass) at $z=0$. It also reproduces observed UV luminosity functions of bright galaxies at $5 \leq z \leq 10$ \cite{Wu24}. However, GRUMPY does not include formation of first Population III stars, which should occur in very low-mass halos of about $10^6$ solar masses.

The GRUMPY model output includes histories of stellar, molecular, and neutral gas masses in galaxies. Ionizing photon production rates were calculated using the BPASS model \cite{Eldridge17,Byrne22}, which includes effects of binary stars and alpha elements. We used the default initial mass function from BPASS. Note that BPASS results do not affect star formation rates in our model; feedback from stellar UV radiation was accounted for differently in GRUMPY, using approximations from radiation transfer simulations \cite{Gnedin14}.

After applying GRUMPY and BPASS, each galaxy's Lyman continuum photon production rate is obtained. Not all photons escape the galaxy to contribute to intergalactic reionization; only a fraction $f_{esc}$ escapes. This fraction is difficult to compute or measure because it depends on the spatial distribution of dust and gas in galaxies, especially around young stars \cite{Ma15}. Measurements of $f_{esc}$ exist for nearby galaxies \cite{Begley22}, but it is uncertain whether these apply at high redshifts due to complex dust behavior \cite{Nath23,Shchekinov25}. Literature debates $f_{esc}$ magnitude and its dependence on galaxy mass and redshift, see e.g. \cite{Munoz24}. Our results cover $f_{esc}=0.02 - 0.2$, encompassing plausible values.

The required UV photon emission rate per unit volume for reionization is given by \cite{Madau99}:
\begin{equation}
    \dot{n}_{ion} = 10^{49.6} \mathrm{s}^{-1} \mathrm{Mpc}^{-3} \left(\frac{1+z}{6}\right)^3 C,
\end{equation}
where $C \equiv \langle n_H^2 \rangle / \langle n_H \rangle^2$. Numerical models \cite{Pawlik15} show $C$ is 2–5 at the start of reionization, increasing to 10–30 at the end. Quasar absorption observations suggest possibly higher $C=6^{+11}_{-2.4}$ at reionization onset \cite{Davies21}. We consider $C$ between 3 and 10.

Also, according to \cite{Wu24}, on whose methodology we rely, the main contribution to early reionization comes from low-luminosity galaxies, $M_{UV} \geq -14$. This justifies our choice of simulation box size 7 Mpc. Such a cube lacks rare bright galaxies, but their presence is not required to track reionization onset. Meanwhile, the small box size allows high resolution and modeling of low-luminosity galaxies.

\section{Results}
\noindent
For all models, UV photon production rates per volume were calculated for redshifts $z=8-25$. They are described with better than 30\% accuracy by
\begin{equation}
    \log_{10} \dot{n}_{ion}(z) [\mathrm{s}^{-1}\mathrm{Mpc}^{-3}] = a \log_{10}(t + b) + c + \log_{10} f_{esc},
    \label{eq:fits}
\end{equation}
where $t$ is the Universe age in Gyr, and parameters $a,b,c$ are given in Table~\ref{tab:fit_params}.

\begin{table}[t]
    \centering
    \begin{tabular}{l|c|c|c} \hline\hline
    Model & $a$ & $b$ & $c$ \\ \hline
    $\Lambda$CDM & 4.69 & -0.047 & 53.3 \\
    Bump $k_0=4.7$ Mpc$^{-1}$ & 3.0 & -0.105 & 53.8 \\
    Bump $k_0=10$ Mpc$^{-1}$ & 2.79 & -0.078 & 53.4 \\
    Bump $k_0=20$ Mpc$^{-1}$ & 2.33 & -0.078 & 52.8 \\
    Bump $k_0=34$ Mpc$^{-1}$ & 2.95 & -0.065 & 52.7 \\
    Bump $k_0=54$ Mpc$^{-1}$ & 8.09 & 0.256 & 52.6 \\
    Tilt & 0.95 & -0.028 & 52.1 \\ \hline
    \end{tabular}
    \caption{Parameters in formula (\ref{eq:fits}) for UV photon emission rate dependence on Universe age for $8 < z < 25$.}
    \label{tab:fit_params}
\end{table}

\begin{figure}
    \centering
    \includegraphics[width=\linewidth]{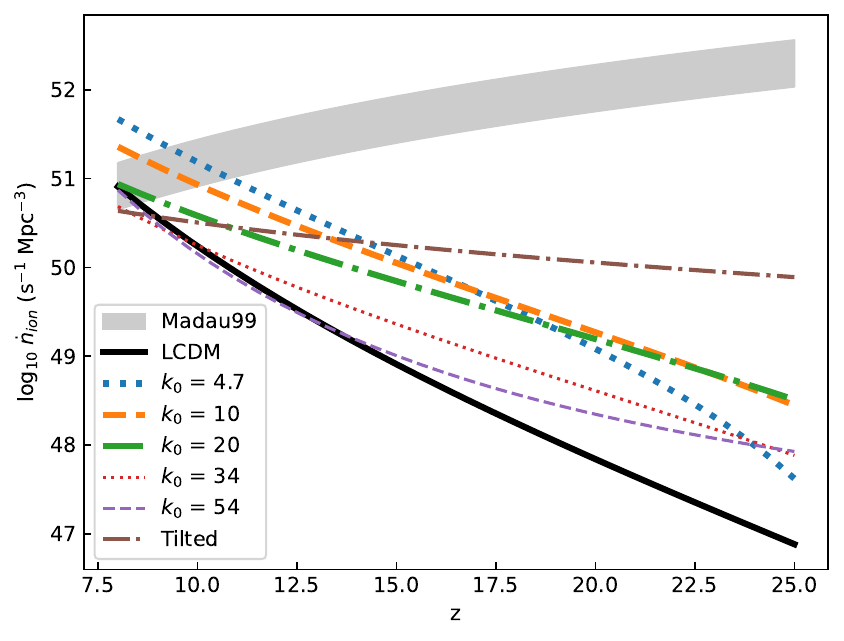}
    \caption{UV photon emission rate. The shaded region shows the photon number required for full ionization for clumping factor $C=3-10$.}
    \label{fig:nion}
\end{figure}

\begin{figure*}
    \centering
    \includegraphics[width=\linewidth]{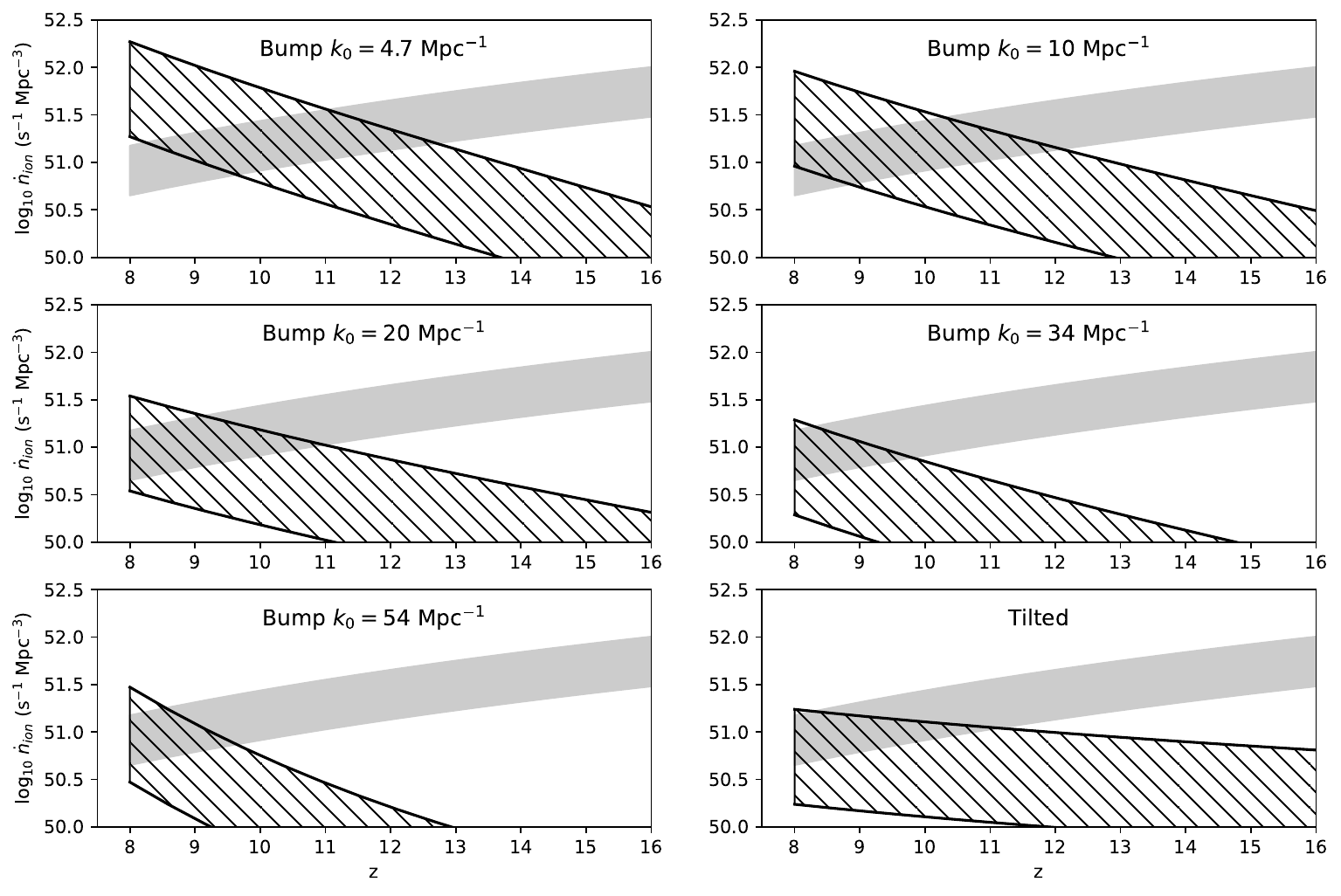}
    \caption{Shaded areas show the range of UV photon emission rates in each distorted spectrum model for escape fractions $f_{esc} = 0.02 - 0.2$. The filled region shows the required photon number for full ionization for clumping factors $C=3-10$. The intersection determines the possible redshift range of reionization.}
    \label{fig:zion}
\end{figure*}

To compare initial spectrum models, we fixed $f_{esc} = 0.05$. The resulting UV photon emission rates vs. redshift are shown in Fig.~\ref{fig:nion}. At $f_{esc}=0.05$, reionization in the standard spectrum model ($\Lambda$CDM) occurs at $z \approx 8$, consistent with \cite{Wu24}. The strongest deviation at $z \sim 10$ from the standard spectrum is in the bump model with $k_0=4.7$ Mpc$^{-1}$. This deviation weakens with increasing $k_0$. At higher redshifts $z \sim 20$, the largest deviation (by 3 orders of magnitude) is in the tilted spectrum model, followed by $k_0=10$ and 20 Mpc$^{-1}$ bump models. Despite large deviations at high $z$, all models remain at least two orders of magnitude below the photon emission rate required for reionization.

To constrain allowable spectrum deviations, we compare models in more detail considering $f_{esc}$ uncertainty in Fig.~\ref{fig:zion}, where each model (except $\Lambda$CDM) is shown on a separate panel. Since $f_{esc}$ and clumping factor $C$ are poorly known, substantial hydrogen ionization can occur at any point in the intersection of the shaded and hatched regions. Only the bump model with $k_0=4.7$ Mpc$^{-1}$ has this intersection at $z > 8$, and even then a slight reduction in bump amplitude, $\mathcal{A}$, or an increase in $C$ can lower this below $z=8$.

All our numerical models were done with bump amplitude $\mathcal{A} = 2$, but we expect roughly linear dependence of the difference in $\dot{n}_{ion}$ between bump and no-bump models on $\mathcal{A} \leq 2$, since the halo mass function depends approximately linearly on this amplitude at such values. To formally satisfy the intersection condition at $z < 8$ for the $k_0=4.7$ Mpc$^{-1}$ model, it suffices to reduce the bump amplitude to $\mathcal{A} = 1.5$.

\section{Conclusion}
\noindent
We calculated the ionizing UV photon production rate at high redshifts, $z > 10$, for several initial cosmological power spectrum models with a bump (\ref{bump}) and tilt (\ref{tilt}), motivated by observations of an increased number of galaxies at high redshifts. Calculations used the semi-analytical GRUMPY model, previously applied by other authors \cite{Wu24} to study reionization. This model was applied to results of numerical dark matter N-body simulations.

Significant uncertainties in the fraction of escaping UV photons $f_{esc}$ and the neutral hydrogen clumping factor $C$ prevent precise determination of reionization timing. Observations of reionization at $z \approx 8$ exclude only bump models with $k_0 \leq 5$ Mpc$^{-1}$ and amplitude $\mathcal{A} > 1.5$. Such models require extremely low escape fractions $f_{esc} < 0.02$ or high clumping factors $C > 10$.

Nonetheless, other models, especially with tilt on small scales or bumps at $k_0 \geq 10$ Mpc$^{-1}$, show significant (orders of magnitude) increases in UV photon production rates compared to the standard spectrum at $z > 15$. Although insufficient to fully ionize the Universe at such high $z$, this may strongly affect 21 cm line absorption of relic photons, as discussed in \cite{Munoz20,Eroshenko25,Naik25}, and 21 cm observations may provide better constraints.

This work by O.R. Skorikov and S.V. Pilipenko is supported by the BASIS Foundation grant for theoretical physics and mathematics development.

\bibliographystyle{ieeetr}
\bibliography{refs}

\end{document}